\documentclass[twocolumn,prl]{revtex4}
\usepackage{float}
\usepackage{makeidx}
\usepackage{graphicx,amsmath}

\begin{document}

\title{Experimental study of balanced optical homodyne and heterodyne detection by controlling sideband modulation}
\author{Wei Li, Zengming Meng, Xudong Yu, Jing Zhang$^{\dagger}$}
\affiliation{State Key Laboratory of Quantum Optics and Quantum
Optics Devices, Institute of Opto-Electronics, Shanxi University,
Taiyuan 030006, P.R.China \label{in}}

\begin{abstract}

We experimentally study optical homodyne and heterodyne detections
with a same setup, which is flexible to manipulate the signal
sideband modulation. When the modulation only generate a single
signal sideband, the light field measurement by mixing the single
sideband at $\omega_{0}+\Omega$ with a strong local oscillator at
the carrier frequency $\omega_{0}$ on a beam splitter become
balanced heterodyne detection. When two signal sidebands at
$\omega_{0}\pm\Omega$ are generated and the relative phase of the
two sidebands is locked, this measurement corresponds to optical
balanced homodyne detection. With this setup, we may confirm
directly that the signal-to-noise ratio with heterodyne detection is
two-fold worse than that with homodyne detection. This work will
have important applications in quantum state measurement and quantum
information.

\end{abstract}

\maketitle
\section{Introduction}
Optical homodyne and heterodyne detections are the extremely useful
and flexible measuring methods and have the broad applications in
optical communication \cite{1,2}, weak signal detection \cite{3}. In
general, optical homodyne and heterodyne detections are defined as
the difference of the optical frequencies of the two mixed field is
zero and nonzero respectively. When the signal is mixed with a
relative strong local oscillator (LO) on a 50$\%$ beam splitter, the
two output modes are detected by a pair of detectors and the
difference of two photocurrents can be measured, this scheme
corresponds to balanced homodyne and heterodyne detections. It is
well known that balanced detection can cancel the noise resulting
from local oscillator \cite{4,5}. Balance homodyne detection is
phase sensitive and has been intensively utilized to measure
non-classical states of light \cite{6,7,8,9,10} and further in
quantum information experiments \cite{11,12,13,14}. Heterodyne
detection has been shown that its signal-to-noise ratio is two-fold
worse than that with homodyne detection \cite{4,5,15,16}.

\section{Experimental setup and results}
In this paper, we use sideband method \cite{17,18} in quantum optics
to investigate balanced homodyne and heterodyne detections in a same
setup. When only mixing the single sideband signal at
$\omega_{0}+\Omega$ with a strong LO at the carrier frequency
$\omega_{0}$ on a 50$\%$ beam splitter, this measurement corresponds
to balanced heterodyne detection. When there are two signal
sidebands at $\omega_{0}\pm\Omega$ and the relative phase of the two
sidebands is locked, this measurement corresponds to balanced
homodyne detection. Although the frequencies of two sidebands are
different with that of LO, it still refer as balanced homodyne
detection since two sidebands locate symmetrically at the two sides
of the carrier frequency of LO, and at the same time the relative
phase of two sidebands is fixed. This balanced homodyne detection
for two sidebands has been exploited widely in quantum optics
experiments \cite{6,7,8,19,20}. We will demonstrate that this
balanced homodyne detection is also phase-sensitive and its
signal-to-noise ratio is two-fold better than that with heterodyne
detection. This work gives the deeper understanding of optical
homodyne and heterodyne detections and the closely relationship
between them. Moreover, the scheme of generating two sidebands in
this paper is different from the general method of phase-modulation
(or amplitude-modulation) by an electro-optic modulator (EOM). A
laser with frequency $\omega_{0}$ is phase-modulated (or
amplitude-modulated) by EOM with frequency $\Omega$ to generate two
sidebands $\omega_{0}\pm\Omega$, hence, this light includes a strong
carrier field and two sidebands. The signal field in this work only
has two sidebands and the optical field at carrier frequency
$\omega_{0}$ is vacuum, which will have the special applications in
quantum information and communication.

\begin{figure}
\centerline{\includegraphics[width=3in]{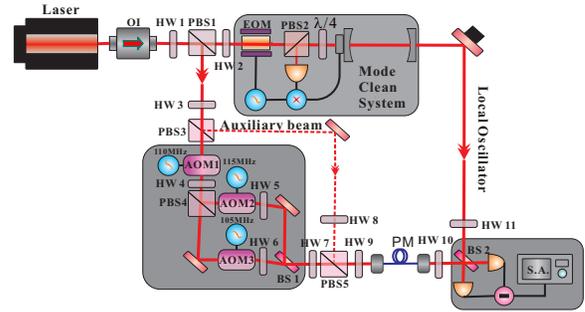}}
\caption{(Color online) Schematic diagram of the experimental setup.
OI: optical isolator. HW: half-wave plate. PBS: polarizing beam
splitter. AOM: acousto-optic modulator. PM: single-mode
polarization-maintaining fiber. EOM: electro-optic modulator. SA:
Spectrum analyzer. }
\end{figure}

Figure 1 shows the experimental setup. A continuous-wave
single-frequency coherent laser at 1064 nm is split into two parts.
One passes through a mode clean cavity to be used for the LO in the
balanced detection system. The other is sent through three
acousto-optic modulators (AOMs, 3110-197, Crystal Technology) for
generating signal sidebands. The laser first is frequency
down-shifted by the negative first-order diffraction of AOM1 with
the frequency -110 MHz. Then the downshift laser is split into two,
which are frequency up-shifted by the positive first-order
diffraction of AOM2 with +115 MHz and AOM3 with +105 MHz,
respectively. The two frequency-shifted beams are combined on 50$\%$
BS1 with the same polarization, so the two sidebands at
$\Omega_{+(-)}=5$ MHz is generated. Here, the sinusoidal signals of
AOMs are provided by three signal generators (N9310A, Agilent)
respectively and the diffraction efficiency of all AOM's are about
70 $\%$. The two-sideband field is coupled into a single-mode
polarization-maintaining fiber and then combined with a strong local
oscillator on 50$\%$ BS2 with the same polarization. The output
fields of BS2 are detected by two balanced detectors, the
substraction of whose photocurrents is measured by the spectrum
analyzer. The single sideband can be obtained by blocking the
incident beam of AOM2 (or AOM3). Here, an auxiliary beam without
frequency shift is used to adjust the visibility of the interference
between the local and the signal. When measuring data, the auxiliary
beam will be blocked.

\begin{figure}
\centering
\includegraphics[width=3in]{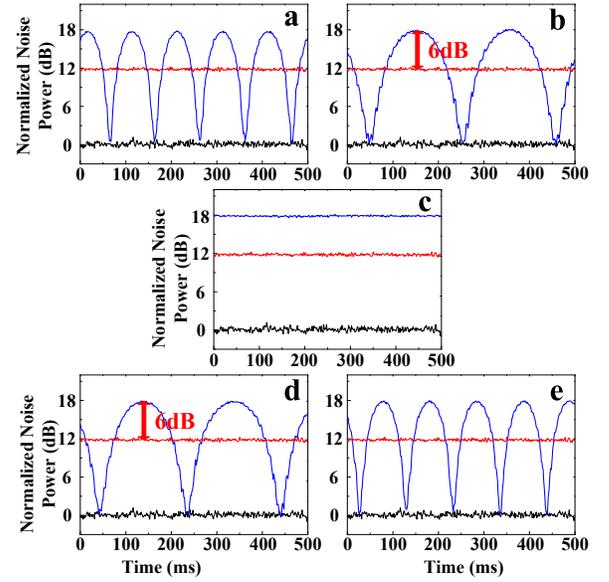}\vspace{0.1in}
\caption{(Color online) The measured noise spectra of the signal
beam by the balanced detection system. The modulation frequency of
AOM1 and AOM2 is fixed at 110 MHz and 115 MHz respectively, which
correspond to up sideband frequency $\Omega_{+}=$5 MHz. The
frequency of AOM3 is changed to set the down sideband frequency
$\Omega_{-}$. The black line is the shot noise limit and the red
line corresponds to the single sideband. The blue lines in (a)-(e)
correspond to the two sidebands cases with the different frequency
difference of up and down sideband
$\Delta\Omega_{\pm}=\Omega_{+}-\Omega_{-}$: -10 Hz, -5 Hz, 0 Hz
(with $\theta=0$), +5 Hz, +10 Hz respectively. The center frequency
of the spectrum analyzer is 5 MHz and the span is zero. RBW is 100
kHz, VBW is 300 Hz and the sweep time is 500 ms. Here, the frequency
difference of up and down sideband $\Delta\Omega_{\pm}$ is much less
than the RBW and VBW. }
\end{figure}

The difference photocurrent of the balanced detection can be
expressed as
\begin{eqnarray}
\delta i&=&i_{1}-i_{2}\nonumber \\
&=&a^{\dagger}_{LO}a_{s}+a_{LO}a^{\dagger}_{s}.\label{BD}
\end{eqnarray}
The LO field can be written as $a_{LO}=\langle a_{LO}\rangle
e^{i(\omega_{0}t+\theta)}$, and the signal field as $a_{s}=\langle
a_{+}\rangle e^{i(\omega_{0}+\Omega_{+})t}+\langle a_{-}\rangle
e^{i(\omega_{0}-\Omega_{-})t}$, which includes two sidebands. Here,
$\theta$ is the relative phase between the LO and signal field.
Hence, the photocurrent of the signal field is written as
\begin{eqnarray}
\delta i_{s}=2\langle a_{LO}\rangle[\langle a_{+}\rangle
\cos(\Omega_{+}t-\theta)+\langle a_{-}\rangle
\cos(\Omega_{-}t+\theta)].\label{Sig}
\end{eqnarray}
Here, two terms represent the two beatnotes of the two sidebands
with LO respectively. Case 1: When only one signal sideband is
applied ($a_{+}$ or $a_{-}$), the signal strength become $(2\langle
a_{LO}\rangle \langle a_{s}\rangle)^{2}$ at the frequency of
$\Omega_{s}$, where $\langle a_{s}\rangle=\langle a_{+(-)}\rangle$
and $\Omega_{s}=\Omega_{+(-)}$. We can see the signal strength
becomes constant and is independent of the relative phase $\theta$
between the LO and signal field for single signal sideband, as shown
in Fig. 2 (red line). This case corresponds to balanced heterodyne
detection. Case 2: When two signal sidebands are applied at the same
time and have the same amplitude with $\langle a_{s}\rangle=\langle
a_{+}\rangle=\langle a_{-}\rangle$, the difference photocurrent can
be written as
\begin{eqnarray}
\delta i_{s}&=&2\langle a_{LO}\rangle\langle a_{s}\rangle
\cos(\frac{\Omega_{+}-\Omega_{-}}{2}t-\theta)
\cos(\frac{\Omega_{+}+\Omega_{-}}{2}t).\label{Sig}
\end{eqnarray}
The signal strength measured by spectrum analyzer becomes
$4(2\langle a_{LO}\rangle \langle
a_{s}\rangle)^{2}\cos^{2}[(\Omega_{+}-\Omega_{-})t/2-\theta]$ at the
frequency of $(\Omega_{+}+\Omega_{-})/2$. Compared with the single
signal sideband, there are several new characteristics (blue line in
Fig. 2 (a), (b), (d) and (e). First, the signal strength has a
periodical modulation with the frequency
$\Delta\Omega_{\pm}=\Omega_{+}-\Omega_{-}$. Second, the maximum
signal strength for two signal sidebands is four-fold larger than
that for single signal sideband. The minimum signal strength reaches
the shot noise limit. When the difference of the frequencies of two
signal sidebands is zero ($\Delta\Omega_{\pm}=0$), the signal
strength becomes $4(2\langle a_{LO}\rangle \langle
a_{s}\rangle)^{2}\cos^{2}(\theta)$. Here, the relative phase
$\theta$ may be the relative phase between up and down sideband or
between the LO and signal field. When the relative phase between up
and down sideband is fixed, this case corresponds to balanced
homodyne detection, which is phase sensitive to the relative phase
$\theta$ between the LO and signal field. When the relative phase
$\theta$ is zero, the signal strength reaches the maximum value
(blue line in Fig. 2(c)), which corresponds the constructive
inference of two sidebands. When the relative phase $\theta$ is
$\pi/2$, the signal strength become zero (shot noise level), which
corresponds the destructive inference of two sidebands.

\begin{figure}
\centering
\includegraphics[width=3in]{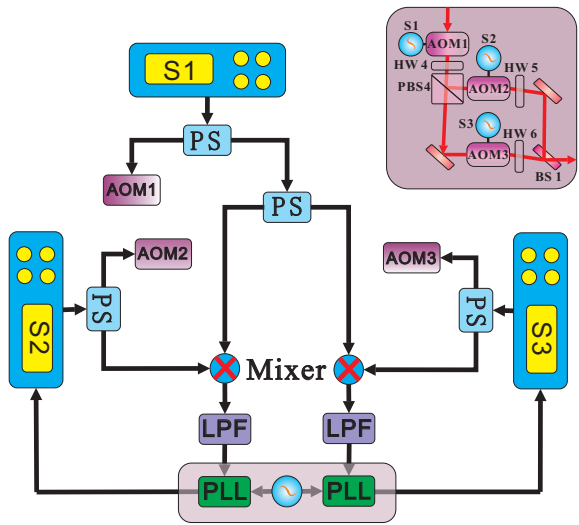}\vspace{0.1in}
\caption{(Color online) The schematic diagram of the method 1 for
locking the relative frequency and phase of the two sidebands. S1-3:
Signal generator. PS: Power splitter. AOM: Acousto-optic modulator.
LPF: Low-pass filter. PLL: phase locked loop. }
\end{figure}
As above discussion, only when the two signal sideband frequencies
$\Omega_{+(-)}$ are same and the relative phase is fixed, the
measurement becomes the balanced homodyne detection. Here we develop
two methods to lock the relative phase between up and down sideband.

\emph{Method 1. }Figure 3 shows the schematic diagram of locking the
relative phase between up and down sideband. The sinusoidal signal
output of the signal generator 1 is divided two parts: one is used
to drive the AOM1 and the other is further divided two parts to mix
with the output of the signal generator 2 and 3. The sinusoidal
signal output of the signal generator 2 (3) is also divided two
parts: one is used to drive the AOM2 (AOM3) and the other is to mix
with the output of the signal generator 1. Two sinusoidal signals
from the signal generators 1 and 2 (1 and 3) are mixed down to
generate about 5 MHz signal using a mixer (Mini-Circuits ZAD-6+).
The mixer's output signal of about 5 MHz pass through a low-pass in
order to filter the high frequency signals of the inputs of the
mixer. Thus we obtain two ways of 5 MHz signal, which correspond to
two sideband frequencies $\Omega_{+}$ and $\Omega_{-}$ respectively.
Two ways of 5 MHz signal are electronically phase-locked to a same 5
MHz reference signal by two phase locked loops (PLL). The output
error signal of the PLL contains the relevant information abut the
frequency and phase difference between two input signals. Two ways
of the error signal, after the proportional-integral-derivative
controller, are feedback into the signal generator 2 and 3
respectively. When the frequency are same and the relative phase
between up and down sideband is locked, the optical field with two
signal sidebands can be phase-sensitively detected by the balanced
homodyne detection.

By controlling the phase of the LO, the relative phase $\theta$
between the LO and signal field can be fixed with zero and the
signal strength reaches the maximum value (blue line in Fig. 5(a)).
When the relative phase $\theta$ is fixed with $\pi/2$, the signal
strength becomes zero and reaches the shot noise level (pink line in
Fig. 5(a)). This pink line presents the large fluctuation due to
imperfect locking. When we scan the phase of LO field, the signal
strength presents a periodical change (green line in Fig. 5(a)).
\begin{figure}
\centering
\includegraphics[width=3in]{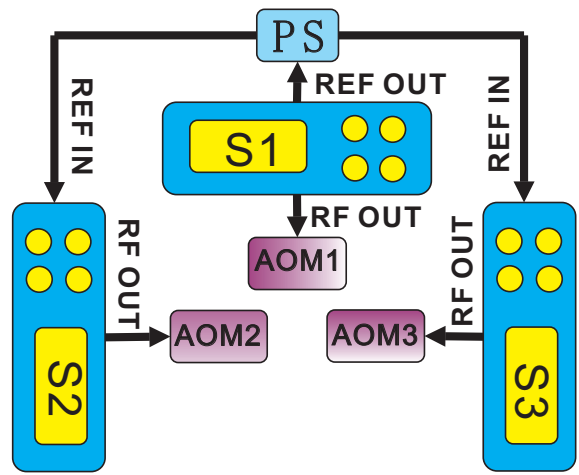}\vspace{0.1in}
\caption{(Color online)  The schematic diagram of the method 2 for
locking the relative frequency and phase of the two sidebands. S1-3:
Signal source. PS: Power Splitter. AOM: Acousto-optic modulator. }
\end{figure}

\emph{Method 2. } Here, we present a simpler scheme to make the
frequencies and the relative phase of the up and down sidebands
locked by using the clock synchronization of Agilent signal
generators. The three signal generators can be locked together in
frequency and phase by using the same reference (clock) frequency.
So the sinusoidal signal outputs are generated from the same clock
source. For instance, the signal generators 1 serves as the master
reference, whose signal first is locked with its internal reference.
The reference output of the signal generators 1 is connected with
the external reference input of the signal generators 2 and 3. The
signals of the signal generators 2 and 3 are locked with the
external reference. The result of the balanced homodyne detection is
shown as Fig. 5(b), which is better than that with the method 1.

\begin{figure}
\centering
\includegraphics[width=3.0in]{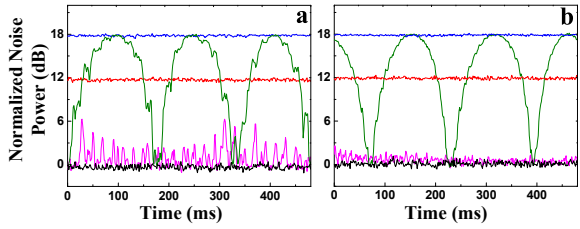}\vspace{0.1in}
\caption{(Color online) The noise spectra measured after locking the
frequency and phase of the two sidebands with two different methods.
(a) use method 1. (b) use method 2. The red curve shows the noise
spectra of single sideband. The blue and purple curves are for the
relative phase $\theta$ between the LO and signal field with zero
and $\pi/2$, respectively. The green curve shows the noise spectra
of two sidebands while scanning the relative phase $\theta$ between
the LO and signal field. RBW = 100 kHz, VBW = 300 Hz, Sweep time=500
ms. }
\end{figure}
We further measure the noise spectra with the frequency span of 3
MHz when locking the frequency and phase of the two sidebands with
method 2. The red line in Fig. 6 presents the case of one signal
sideband, which corresponds to balanced heterodyne detection. The
signal-to-noise ratio for one signal sideband is 12 dB, which is
independent of the relative phase $\theta$ between the LO and signal
field for single signal sideband. When two signal sidebands are
applied and the phase of the two sidebands are locked, the
signal-to-noise ratio is sensitive to the relative phase $\theta$
between the LO and signal field for two signal sidebands, which
corresponds to balanced homodyne detection. The maximum
signal-to-noise ratio for two signal sidebands is four-fold (6 dB)
larger than that for single signal sideband when the relative phase
$\theta=0$. When consider the factor 2 of two sideband signals
compared with single sideband, we confirm that the signal-to-noise
ratio with heterodyne detection is two-fold worse than that with
homodyne detection. The signal-to-noise ratio for two signal
sidebands reaches zero when the relative phase $\theta=\pi/2$.
\begin{figure}
\centering
\includegraphics[width=3.0in]{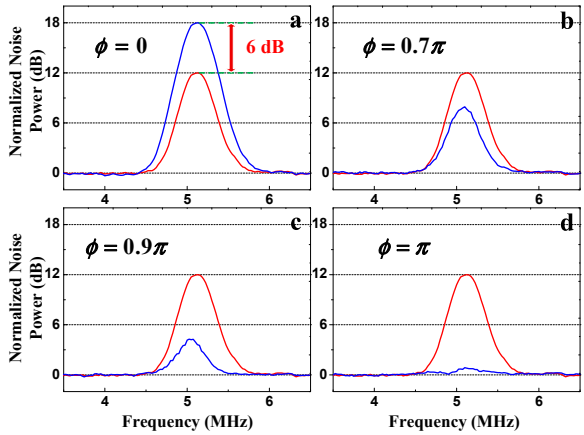}\vspace{0.1in}
\caption{(Color online) The noise spectra with a certain span at the
analyzed frequency when the relative frequency and phase of the two
sidebands are locked. The red curves and the blue curves represent
single sideband and the double sideband cases respectively. The span
of the measured frequency is from 3.5 MHz to 6.5 MHz. RBW = 100 kHz,
VBW = 30 Hz, Sweep time=500 ms. }
\end{figure}
\section{Conclusion}
In conclusion, we experimentally study optical balanced homodyne and
heterodyne detections by the sideband method. The single sideband
and two sideband signals can be obtained easily in our experimental
setup, therefore optical balanced homodyne and heterodyne detections
can be investigated and compared simultaneously. We also develop two
methods to lock the relative phase between up and down sideband for
realizing the balanced homodyne detection. We confirm that this
balanced homodyne detection scheme is phase-sensitive and its
signal-to-noise ratio is two-fold better than that with heterodyne
detection. The scheme of generating single and two sideband signal
in this work can be used in the quantum information and quantum
metrology \cite{21,22,23}. Resently, we noticed an interesting work
where the heterodyne detection with a bichromatic local oscillator
was theoretically and experimentally studied \cite{24}.

$^{\dagger}$Corresponding author email: jzhang74@aliyun.com,
jzhang74@sxu.edu.cn

\begin{acknowledgments}
This research was supported in part by National Basic Research
Program of China (Grant No. 2011CB921601), NSFC for Distinguished
Young Scholars (Grant No. 10725416), NSFC Project for Excellent
Research Team (Grant No. 60821004).
\end{acknowledgments}

\end{document}